# HIERARCHICAL TAXONOMY IN MULTI-PARTY SYSTEM

Hokky Situngkir[*)]
(hokky@elka.ee.itb.ac.id)
Dept. Computational Sociology
Bandung Fe Institute

**Abstract**

We propose the use of hierarchical taxonomy to analyze the legislative election results as a model of multi-party system to show the robustness in political system. As an example we use the result of Indonesian legislative election 2004 is analyzed with certain comparative with the previous one (1999). We construct the graph theoretical analysis by finding the Euclidean distances among political parties. The distances are then treated in ultrametric spaces by using the minimum spanning tree algorithm. By having the Indonesian hierarchical taxonomy model of political parties we show some patterns emerging the pattern agrees with the classical anthropological analysis of socio-political system in Indonesia. This fact accentuates a character of robustness in Indonesian political society as a self-organized system evolves to critical state. Some small perturbations i.e.: different voting process resulting the same pattern and occasions statistically, emerges from the social structure based upon political streams: Islamic, secular, traditional, and some complements of all.

## 1. Introduction

In the previous work (Situngkir, 2004) we have showed how the social system evolves toward the critical self-organization by analyzing the statistical properties of the national elections in 1999 and 2004. The power-law signature of the general elections will be tried to be analyzed deeper by analyzing the political structures of the voters in Indonesia; why and how it occurs. There are some patterns, and the paper aims to explore some scale-invariant causes of the election results. We use method that has been used more familiarly in econophysics, i.e.: the Euclidean distances among parties based on their votes and find to describe the statistical situations into the ultrametric spaces by using the minimum spanning tree algorithm. Eventually, we will find out portfolio-like diagram evolves from 1999 to 2004 elections and that there happens the political robustness in the political structure of the voters (Mantegna & Stanley, 2000:105-12).

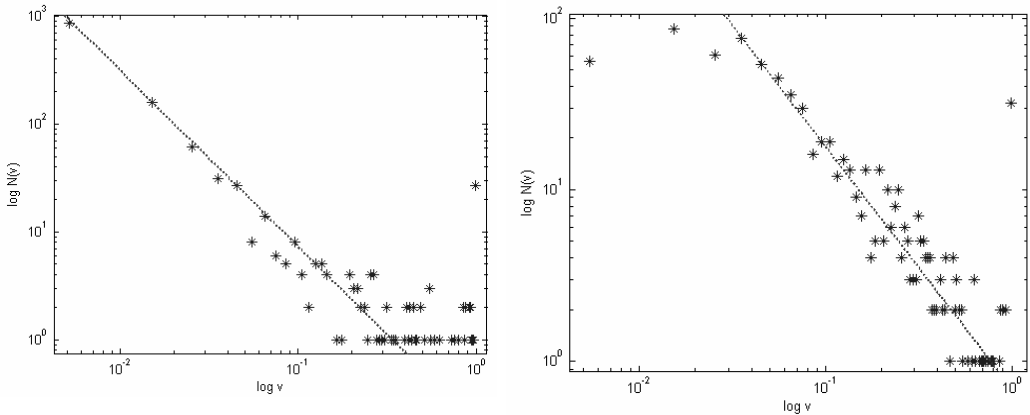

**Figure 1**
The power-law signature of Indonesian legislative election result 1999 and 2004.

## 2. Visualizing political streams in ultra-metric space

We normalize the votes of each political party by the highest vote in each province from data of the election result held in 1999 and 2004. As it has been analyzed in Situngkir (2004), we built the histogram with a unique histogram showing the number of political parties, $N(v)$, that received certain fraction of votes, $v$. The log-log plot of both histograms show power-law signature as figured in figure 1. The election result shows that the data is fitted with power-law distribution, $N(v) \sim v^{-\alpha}$, $\alpha = 1.632$ and $\alpha = 1.41$, for 1999 and 2004 election respectively.

After the normalization, we find the cross-correlation among the party in each province to construct cross-correlation matrix. that shows the cross-correlation coefficient among party $i$ with party $j$,

$$\rho_{ij} = \frac{\langle V_i V_j \rangle - \langle V_i \rangle \langle V_j \rangle}{\sqrt{\langle V_i^2 - \langle V_i \rangle^2 \rangle \langle V_j^2 - \langle V_j \rangle^2 \rangle}} \qquad (1)$$

$V_i$ is the normalized votes of party $i$ in each province and the angular brackets indicate the average of the votes. Here, we can have the correlation coefficient $\rho_{ij} = [-1..1]$, where

$$\rho_{ij} = \begin{cases} 1, \text{ completely correlated} \\ 0, \text{ uncorrelated} \\ -1, \text{ completely anti correlated} \end{cases} \qquad (2)$$

We use the cross-correlation coefficient among parties by modificating the calculation of Euclidean distance among log-price difference (Mantegna & Stanley, 2000: 105-6) to extract the information hiding in the election result by calculating the distances among parties. By constructing the algebraic vectors on closely related parties, it was found out that the Euclidean distances ($d_{ij}$) among parties can be calculated as

$$d_{ij} = \sqrt{2(1 - \rho_{ij})} \qquad (3)$$

Fulfilling the properties of Euclidean distance, it should be held by properties

$$properties \rightarrow \begin{cases} d_{ij} = 0 \Leftrightarrow i = j \\ d_{ij} = d_{ji} \\ d_{ij} \leq d_{ik} + d_{kj} \end{cases} \qquad (4)$$

As explained in (Mantegna & Stanley, 2000: 107, & Mantegna, 1999), in order to construct a hierarchical model of such complex system whose elements have Euclidean distances, we need to form taxonomy about the topological space of *n* objects. We do this by implicating the Minimum Spanning Tree (MST) Algorithm to the *n x n* distance matrix. MST is a concept of weighted graph of *n* objects in which a tree having *n-1* edges that minimize the sum of the edge distances. To do this, we resemble the distance matrix by applying the well-known Kruskal's algorithm. In this case, we can define tree as a connected graph without cycles with some properties, i.e.: there is one and only one path joining any two of its vertices, and that every one of its edges is a bridge. Thus, the MST is a weighted and connected graph having one (possibly more) the least total weight.

The MST technique aims to quantify spatial dot patterns by revealing hidden nearest-neighbor correlations, and the Kruskal's algorithm we use can be stated as follows:

```
begin

    do while (all vertex in the graph)
        Find the least edge in the graph;
        Mark it with any given color, e.g.: blue;
        Find the least unmarked (uncolored) edge in the graph
        that doesn't close a colored or blue circuit;
        Mark this edge red;
    end;

The blue edges form the desired minimum spanning tree;

end
```

The resulting matrix is specially known as description of political parties in an ultrametric space by physicists (Rammal, et. al., 1986).

In summary, the metric of the set of political parties $V$ is given by the assignment of real number $d_{ij}^{ult}$, where $d_{ij}^{ult}$ fulfils requirements of Euclidean properties (eq. 4) with additional requirement, i.e.:

$$d_{ij}^{ult} \leq \max(d_{ik}^{ult}, d_{kj}^{ult}) \qquad (5)$$

By using the algorithm above, we can have the matrix describing the relative closest parties with their neighborhoods for each year of election, 1999 and 2004. The result will be elaborated in the next section.

### 3. The result and analysis

We have the MST of the distances of Indonesian political parties as the result of election in 1999 and 2004, and discover how the parties clustered with certain patterns in the graph. Figure 1 shows the result of our simulation on the data of the General Election 1999. There have been 48 political parties joined the election where the citizens voted for depend on their appropriateness. It is believed that this is the first most democratic election Indonesia ever had after escaping from 32 years of dictatorship regime. The distance scale in the figure equals to the maximal distance between two successive political parties encountered when moving from a certain political party to the other over the shortest path of the MST connecting them (the number of the respective political party can be seen in the appendices).

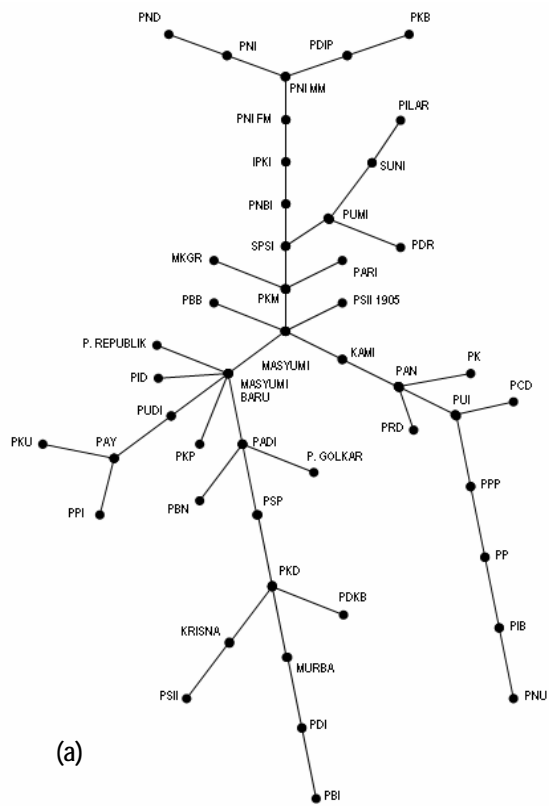

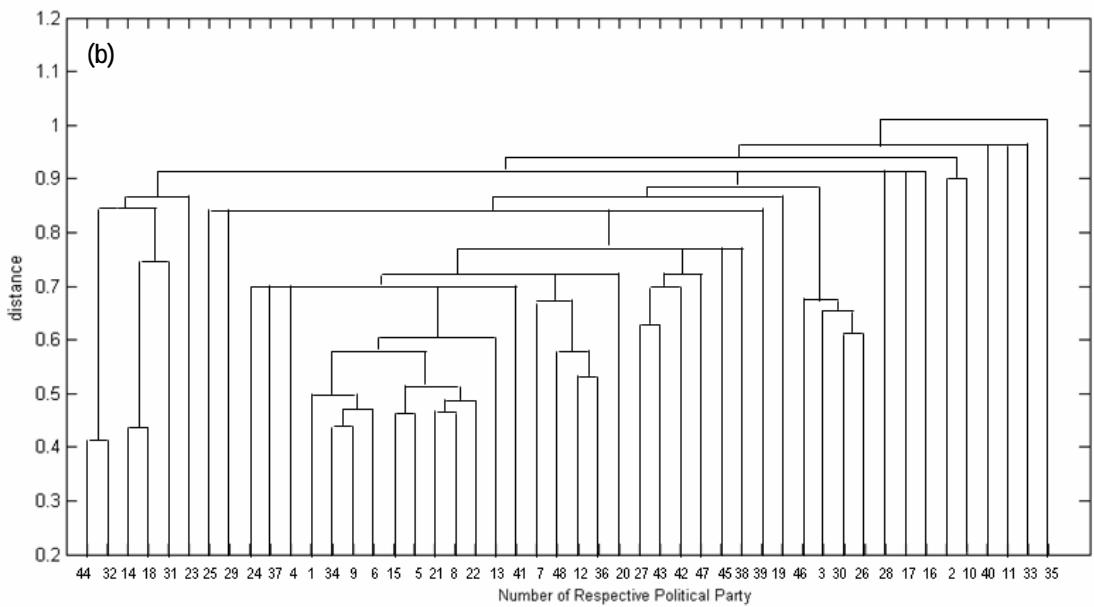

Figure 2
The structure of the MST of the result of the General Election 1999 *(a)*, and The distances among parties in ultrametric space *(b)*

The taxonomy presented here is associated with the subdominant ultrametric of certain political parties. It is a meaningful political tool since we can see how the structure of the gains of political parties regarding their voters. From the figure, we can see how most parties are closed enough while some other parties are separated apart. It is also obvious that some parties are very close to each other since there is no major difference among them perceived by the voters.

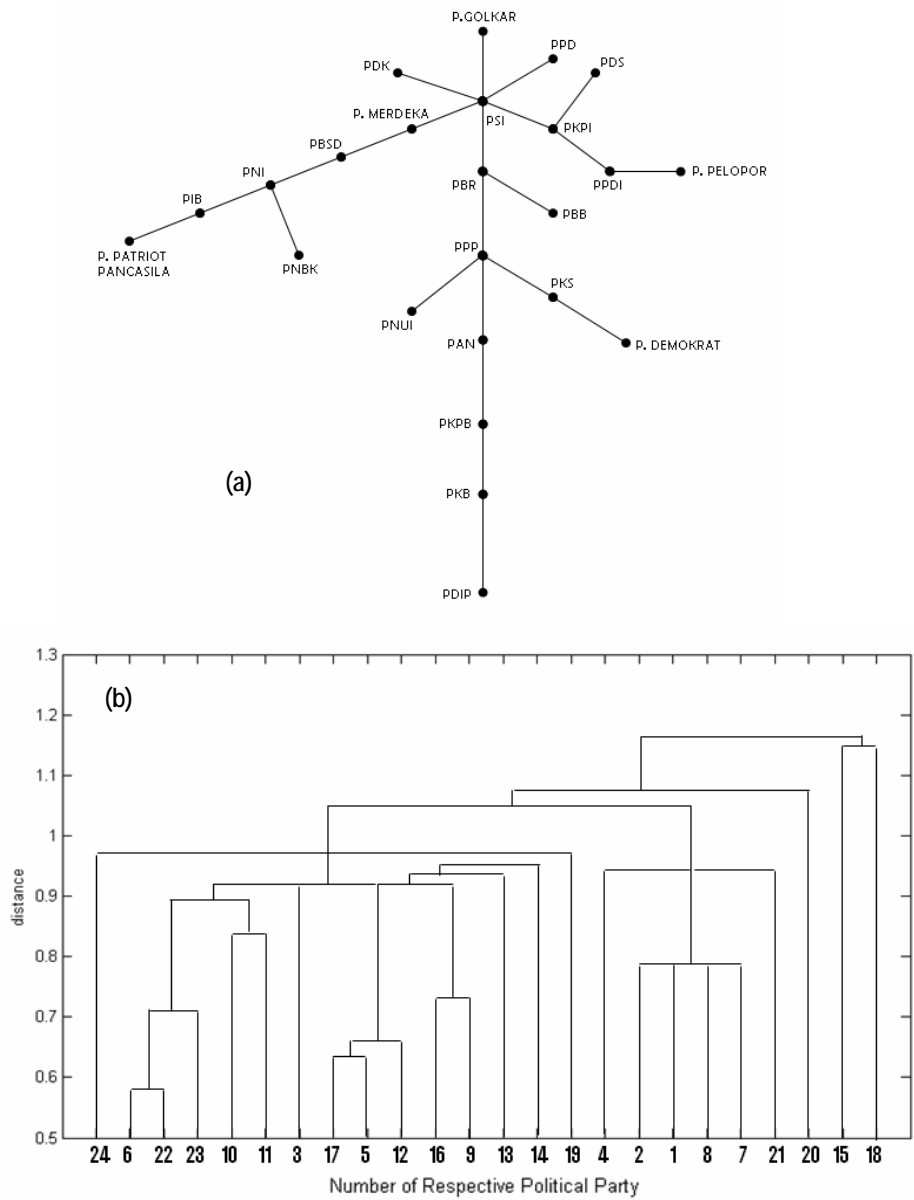

Figure 3
The structure of the MST of the result of the General Election 2004 *(a),* and The distances among parties in ultrametric space *(b)*

Figure 3 shows the MST for the statistics of legislative election 2004. In 2004, the political reform has urged to change the rule of political system including the election system. The political system has changed to be the bicameralism system while General Election is held in order to let citizens choose directly the members of DPR (House of Representatives), DPRD I & II (city councils) and DPD (the Regional Representative Council). Directly is by means of choose not only from the collection of political parties but choose directly the individuals to be seated in certain political institutions. It is obvious that the election 2004 become a hope for a better political system in Indonesia.

Interestingly, compared to the previous one, the legislative election 2004 holds the similar pattern, i.e.: three big parties are very distant (figure 2 & 3). As showed in previous sections, most of other parties are trying to gain votes from the highly networked social institutions circling related political parties. Intuitively, based upon this we can realize how the power-law signature appears in Indonesian general election, since only several parties dominate in several political streams and groups among plenty of available political parties.

## 6. Concluding Remarks

We have showed how methodologically we can have the hierarchical taxonomy of political parties in ultrametric space as a meaningful way to see the clustering of the parties. This can be an alternative on extracting the result of general election across the country as an important statistical property.

Eventually, we show also that the political streams figured out by the hierarchical taxonomy accentuate and is directly related to the anthropological analysis proposed many years before the election. Even further, the significant changes in the micro-stages of the election do not impact directly with the taxonomy – a signature of robustness of socio-political environment.

As an epilogue, several questions are left to the reader about the process reform and democratization in Indonesia. How robust are the circling social organizations relating to the demand of social transformations? If the social networks and social identities have huge influence on deciding the future of a nation, how can this concern with the political program brought by the candidates? It is for sure that we are now having a signal and symptom, that the work and struggle for democracy is about to begin by the democratic process and it is still a long journey to finish line.

**Acknowledgement**

The author thanks Surya Research Inc. for financial support, Tiktik Dewi Sartika, Deni Khanafiah, and Ivan Mulianta for gathering and processing data, Yun Hariadi for deep discussions. All faults remain the author's.

# **APPENDIX 1**

The name and the number of political party in Indonesian General Election 1999

| | |
|---|---|
| 1 | PIB |
| 2 | KRISNA |
| 3 | PNI |
| 4 | PADI |
| 5 | KAMI |
| 6 | PUI |
| 7 | PKU |
| 8 | MASYUMI BARU |
| 9 | PPP |
| 10 | PSII |
| 11 | PDI PERJUANGAN |
| 12 | PAY |
| 13 | PKM |
| 14 | PDKB |
| 15 | PAN |
| 16 | PRD |
| 17 | PSII 1905 |
| 18 | PKD |
| 19 | PILAR |
| 20 | PARI |
| 21 | MASYUMI |
| 22 | PBB |
| 23 | PSP |
| 24 | PK |
| 25 | PNU |
| 26 | PNI FM |
| 27 | IPKI |
| 28 | P. REPUBLIK |
| 29 | PID |
| 30 | PNI MM |
| 31 | MURBA |
| 32 | PDI |
| 33 | GOLKAR |
| 34 | PP |
| 35 | PKB |
| 36 | PUDI |
| 37 | PBN |
| 38 | MKGR |
| 39 | PDR |
| 40 | PCD |
| 41 | PKP |
| 42 | SPSI |
| 43 | PNBI |
| 44 | PBI |
| 45 | SUNI |
| 46 | PND |
| 47 | PUMI |
| 48 | PPI |

**APPENDIX 2**

The name and the number of political party in Indonesian Legislative General Election 2004

| 1 | PNI |
| --- | --- |
| 2 | PBSD |
| 3 | PBB |
| 4 | P.MERDEKA |
| 5 | PPP |
| 6 | PDK |
| 7 | PIB |
| 8 | PNBK |
| 9 | P.DEMOKRAT |
| 10 | PKPI |
| 11 | PPDI |
| 12 | PNUI |
| 13 | PAN |
| 14 | PKPB |
| 15 | PKB |
| 16 | PKS |
| 17 | PBR |
| 18 | PDIP |
| 19 | PDS |
| 20 | P.GOLKAR |
| 21 | P.PAT.PANCASILA |
| 22 | PSI |
| 23 | P.PERS.DAERAH |
| 24 | P. PELOPOR |